\documentclass{article}%
\usepackage{amsmath}
\usepackage{amsfonts}
\usepackage{amssymb}
\usepackage{graphicx}%
\providecommand{\U}[1]{\protect\rule{.1in}{.1in}}
\begin{document}

\title{Relativistic fluid spheres with particular application in cosmology and
gravitational collapse}
\author{Ron Wiltshire\\The Division of Mathematics \& Statistics,\\The University of Glamorgan,\\Pontypridd CF37 1DL, UK\\email: rjwiltsh@glam.ac.uk}
\maketitle

\begin{abstract}
A spherically symmetric comoving fluid solution of Einstein's equations is
adapted for cosmological application by extending the geometry of standard FRW
cosmology using a generalised curvature term. The resulting model retains many
of the known cosmological properties including homogeneity of energy density,
its relationship with internal pressure including equations of state, although
in each case they have a generalised structure. It is shown that the adapted
model does not require the inclusion of the arbitrary cosmological constant
and the vacuum energy solution is discussed in its absence. The Hubble
constant and deceleration parameter are also shown to have a form which
characterises the modified geometry of the new model. These forms are
calculated using current observational data and show how the standard
cosmological geometry can be amended in a way which is consistent with an
observed flat curvature and a decelerating universe. Finally the solution is
also considered in the context of gravitational collapse where it is shown how
fluids spheres obeying a central equation of state can be matched to empty spacetime.

\textbf{Short title: }Relativistic fluid spheres applied to cosmology and
gravitational collapse


\medskip PACs numbers: 0420, 0440, 9880

\end{abstract}

\section{Introduction}

Spherically symmetric fluid solutions of Einstein's equations have been
discussed in depth by many authors for many reasons which include their
important wide ranging applications to cosmology and astrophysics. General
solutions have often been obtained using a metric in isotropic form together
with a comoving system of coordinates. From the point of view this paper the
approach originally described by Kustaanheimo and Qvist \cite{ku}, has been
invaluable although although there was an earlier particular solution
presented by McVittie \cite{MCV33}. Subsequently there have other
investigations which often involved Lie symmetry approaches for example
Stephani \cite{sti}, Stephani, Wolf \cite{stw} and Wiltshire \cite{wiltshire1}%
. In addition have been many comprehensive studies of their mathematical,
physical properties and their various interrelationships by for example Nariai
\cite{nr} ,Chakravarty \cite{ch}, McVittie \cite{mc1966,mc}, Srivastava
\cite{sr}, Sussman \cite{sus,s1988a,s1988b}, McVittie \cite{mc1966}, Knutsen
\cite{kn1986}. Furthermore the host of solutions have \ been summarised by
Krasinski \cite{kras} and also Stephani \textit{et al} \cite{stepbook}.

Nonetheless there remain many outstanding problems associated with the
physical applicability of these solutions. For example in cosmology the
simplest spherically symmetric solution gives rise to the Friedmann,
Robertson, Walker (FRW) standard cosmological models which have recently
required significant adaptation to describe emerging observational data. The
evidence of \ very early epoch, inflation, the notion of vacuum energy and a
negative deceleration parameter in a more or less flat universe (based upon
type 1a supernova surveys Perlmutter, Riess, Schmidt, Garnavich \& coworkers
\cite{perl1, riess1, schmidt, garnavich, perl2}) have created many challenges
for the standard model and general relativity. This has required the
reintroduction of the cosmological constant and the resulting concordance or
$\Lambda CDM$ \ model to reconcile observation with theory by means of large
proportions of dark energy $\Omega_{\Lambda}=0.7\ $and cold dark matter
$\Omega_{M}=0.23$. Recent developments Turner \& Reiss \cite{turner}, Virey
\cite{vir} show that even this may not be sufficient, in fact a bimodal model
may be required to describe past deceleration as well as recent acceleration
of the universe. In the following it will be seen that there is a
generalisation of FRW cosmology which overcomes many of these difficulties
without the introduction of a cosmological constant.

Moreover this \ solution will be shown to simplify \ many difficulties in the
description of problems of gravitational collapse where in practice it is
extremely difficult in practice to match a fluid sphere to Schwarzschild empty
space-time even though the theory is well known, Bonnor \& Vickers \cite{bon}.
This is especially true in cases where it is necessary to incorporate equation
of state between pressure and energy density $p=p(\rho)$ for example at the
centre of the sphere. For this reason the Oppenheimer Snyder model \cite{opp}
essentially a zero pressure FRW solution has often been employed in matching
problems as has been the case in discussions of a potential source for
gravitational waves Babak \& Glampedakis \cite{stan} or a matching problem in
first order rotation Kegeles \cite{kegeles}. \ 

This paper is organised in the following way. Einstein's equations in the
context of general spherically symmetric fluid spheres are introduced in
section 1 with a focus on the particular solution to be discussed. The basic
properties of the solution are presented in section 3 in relation to the FRW
cosmological models and the cosmological constant. The solution is then
discussed in the context of vacuum energy in section 4 whilst section 5
focusses on equations of state and problems of gravitational collapse. Finally
in section 6 the solution is discussed in terms of the Hubble constant,
deceleration parameter and some current data.

\section{Fluid spheres, isotropy, preliminary equations}

It is the intention here to consider the isotropic coordinate system for
which\qquad%
\begin{equation}
ds^{2}=e^{2\lambda}dt^{2}-e^{2\mu}\left(  dr^{2}+r^{2}d\Omega^{2}\right)
\qquad\qquad d\Omega^{2}=d\theta^{2}+\sin^{2}\left(  \theta\right)  d\phi^{2}
\label{metric}%
\end{equation}
where $\lambda=\lambda\left(  r,t\right)  $, $\mu=\mu\left(  r,t\right)  $. In
addition Einstein's field equations will be taken to be%
\begin{equation}
G_{k}^{i}=-8\pi T_{k}^{i}\qquad\qquad T_{k}^{i}=\left(  \rho+p\right)
u^{i}u_{k}-\delta_{k}^{i}p \label{pl59}%
\end{equation}
where $G_{k}^{i}$ is \ the Einstein tensor and $T_{k}^{i}$ is the energy
momentum tensor of the fluid sphere. A comoving coordinate system will assumed
so that the components $u^{i}$ of the velocity four vector satisfy $u^{i}%
u_{i}=1$ with $u^{1}=0$ and $u^{2}=0=u^{3}$. The pressure $p$, energy density
$\rho$ and mass function $m$, see for example, Misner and Sharp \cite{mi},
Cahill and McVittie \cite{ca} may be calculated using:%
\begin{equation}
8\pi p=G_{2}^{2}\qquad8\pi\rho=-G_{4}^{4}\qquad m=\frac{re^{\mu}}{2}\left\{
1+e^{2\left(  \mu-\lambda\right)  }r^{2}\mu_{t}^{2}-\left(  1+r\mu_{r}\right)
^{2}\right\}  \label{is40}%
\end{equation}
where the suffix $r$, $t$ indicates a partial derivative. $\ $In this system
Einstein's field equations satisfy the isotropy condition in the form%
\begin{equation}
G_{1}^{4}=0=G_{4}^{1}\qquad\qquad G_{2}^{2}-G_{1}^{1}=0 \label{isot8}%
\end{equation}
and so%
\begin{equation}
\mu_{rt}-\lambda_{r}\mu_{t}=0 \label{k3}%
\end{equation}
and%
\begin{equation}
\mu_{rr}+\lambda_{rr}+\lambda_{r}^{2}-\mu_{r}^{2}-2\lambda_{r}\mu_{r}%
-\dfrac{\left(  \mu_{r}+\lambda_{r}\right)  }{r}=0 \label{k1}%
\end{equation}
Using the approach of Kustaanheimo and Qvist \cite{ku} the solutions of
(\ref{k3}) and (\ref{k1}) may be expressed in terms of the function
$L=L\left(  x,t\right)  $ where%
\begin{equation}
L=e^{-\mu}\qquad x=r^{2}\qquad e^{\lambda}=A\left(  t\right)  \mu
_{t}=-A\left(  t\right)  \frac{L_{t}}{L} \label{kus49}%
\end{equation}
In this notation (\ref{k3}) and (\ref{k1}) are simultaneously satisfied by
\begin{equation}
L_{xx}=L^{2}F\left(  x\right)  \label{kus51}%
\end{equation}

In general this equation has been studied for the mathematically tractable
cases $F\left(  x\right)  =(ax^{2}+bx+c)^{-5/2}$ often leading to solutions
involving elliptic functions, see for example Stephani \textit{et al}
\cite{stepbook} whilst the much simpler form%

\begin{equation}
L\left(  x,t\right)  =f(t)x+g(t)\qquad\qquad F\left(  x\right)
=0\label{revision}%
\end{equation}
leads to the FRW cosmological models and also the Oppenheimer- Synder model
for gravitational collapse \cite{opp}. However the FRW solution%
\begin{equation}
L\left(  x,t\right)  =\frac{1}{R\left(  t\right)  }\left(  1+\frac{kx}%
{4}\right)  \qquad\qquad F\left(  x\right)  =0\label{rws}%
\end{equation}
is only a subcase of the more general class given by (\ref{revision}) which
will be the primary subject of study below. It will be shown how this
generalised class leads to a new class of fluid spheres possessing uniform
energy density with the prospect of extending the class of FRW cosmologies and
also fluid sources for gravitational collapse.

\section{The solution and basic properties}

In the context here it helpful to defines the function $S\left(  r,t\right)  $
for which%
\begin{equation}
S=\frac{rR}{1+\dfrac{\sigma R^{2}r^{2}}{4}} \label{ss}%
\end{equation}
and a solution of Einstein's equations having the general structure of
(\ref{revision}) in the form:%
\begin{equation}
e^{-\mu}\equiv\frac{r}{S}=\frac{\left(  1+\dfrac{\sigma R^{2}r^{2}}{4}\right)
}{R} \label{mu}%
\end{equation}
so that by (\ref{kus49})%
\begin{equation}
e^{\lambda}=\frac{S_{t}}{S}a^{-\frac{1}{2}}=\left(  \frac{R_{t}}{R}%
-\frac{r^{2}R\left(  2\sigma R_{t}+\sigma_{t}R\right)  }{4\left(
1+\dfrac{\sigma R^{2}r^{2}}{4}\right)  }\right)  a^{-\frac{1}{2}} \label{lam}%
\end{equation}
where $R=R\left(  t\right)  $, $\sigma=\sigma\left(  t\right)  $ and
$a=a\left(  t\right)  $ and satisfy (\ref{k3}) and (\ref{k1}) by direct
substitution. The Ricci scalar curvature for the 3 dimensional space in
(\ref{metric}) with (\ref{mu}) has the value $6\sigma$ so that the solution is
closed when $\sigma>0$, flat when $\sigma=0$ or open when \ $\sigma<0$.

Also note that the FRW solution is obtained as a subcase of (\ref{metric})
with (\ref{mu}) and (\ref{lam}) by writing%
\begin{equation}
\sigma=\frac{k}{R^{2}}\qquad\qquad a=\frac{S_{t}^{2}}{S^{2}}=\frac{R_{t}^{2}%
}{R^{2}}\label{rw}%
\end{equation}
The energy density $\rho=\rho\left(  t\right)  $ from (\ref{is40}) is:%
\begin{equation}
8\pi\rho=3\left(  a+\sigma\right)  \label{rho5}%
\end{equation}
It is a purely time dependent equation and is a generalisation of the normal
Friedmann equation when the condition (\ref{rw}) is also to be included.
Equation (\ref{rho5}) demonstrates that the energy density is a composite of
two functions $a\left(  t\right)  $ and $\sigma\left(  t\right)  $ neither of
which have any \textit{a priori} dependence of the scale factor $R\left(
t\right)  $. However the pressure $p=p\left(  r,t\right)  $ calculated using
(\ref{is40}) is in general a function of both $r$ and $R\left(  t\right)  $
and is given through%
\begin{equation}
8\pi\left(  p+\rho\right)  =\frac{\left(  a_{t}+\sigma_{t}\right)  R\left(
4+\sigma r^{2}R^{2}\right)  }{\left(  r^{2}\sigma R^{2}R_{t}-4R_{t}%
+r^{2}\sigma_{t}R^{3}\right)  }\label{prho}%
\end{equation}

Equations (\ref{rho5}) and (\ref{prho}) represent the generalised form of the
two Friedmann equations which form the basis of standard cosmology. Subsequent
analysis in this paper will be based upon these generalised forms. It can be
shown that (\ref{prho}) may be cast into a more familiar form to cosmologists
by combining it with $S\left(  t\right)  $ defined in (\ref{ss}) and the
derivative of (\ref{rho5}) to give%

\begin{equation}
p\frac{\partial S^{3}}{\partial t}=-\frac{\partial\left(  \rho S^{3}\right)
}{\partial t}\,\qquad\Longrightarrow\qquad\rho_{t}=-\frac{3\left(
p+\rho\right)  S_{t}}{S} \label{state}%
\end{equation}
Notice that when either $r=0$ or alternatively $\sigma=k/R^{2}$ this reduces
to standard form%
\begin{equation}
\rho_{t}=-\frac{3\left(  p+\rho\right)  R_{t}}{R} \label{state5}%
\end{equation}
In the addition the mass function has the form%
\begin{equation}
m=\frac{4\pi\rho S^{3}}{3}=\frac{4\pi\rho\left(  Rr\right)  ^{3}}{3}\left(
1+\dfrac{\sigma R^{2}r^{2}}{4}\right)  ^{-3} \label{mass5}%
\end{equation}
also part of standard cosmology and which also be used in cases of
gravitational collapse to define a fluid sphere boundary.

As can be seen from equation (\ref{lam}) it is not possible in general to
choose a function $a\left(  t\right)  $ such that $e^{\lambda}=1$ for all
values of $r$ and so define a comprehensive proper time variable, that is with
the exception of the FRW cases given by (\ref{rw}). However it is possible for
an observer at $r=\kappa$ to define a \textit{local} proper time variable by
setting%
\begin{equation}
S\left(  \kappa,t\right)  =K\left(  t\right)  \qquad\Longrightarrow\qquad
S\left(  0,t\right)  =R\left(  t\right)  \label{k15}%
\end{equation}
and using (\ref{lam})%
\begin{equation}
a\left(  t\right)  =\frac{K_{t}^{2}}{K^{2}}=\left(  \frac{R_{t}}{R}%
-\frac{\kappa^{2}R\left(  2\sigma R_{t}+\sigma_{t}R\right)  }{4\left(
1+\dfrac{\sigma R^{2}k^{2}}{4}\right)  }\right)  ^{2} \label{k21}%
\end{equation}
In this way (\ref{rho5}) becomes%
\begin{equation}
8\pi\rho=3\left(  \frac{K_{t}^{2}}{K^{2}}+\sigma\right)  \label{den9}%
\end{equation}
whilst the pressure from (\ref{state}) now satisfies%

\begin{equation}
p\frac{dK^{3}}{dt}=-\frac{d\left(  \rho K^{3}\right)  }{dt}\qquad
or\qquad\qquad\rho_{t}=-\frac{3\left(  p+\rho\right)  K_{t}}{K} \label{cos10}%
\end{equation}

Without loss of generality in the following it will be assumed that local
propertime is defined at the coordinate centre $r=0$ so that:%
\begin{equation}
8\pi\rho=3\left(  \frac{R_{t}^{2}}{R^{2}}+\sigma\right)  \label{den19}%
\end{equation}
whilst from (\ref{cos10})%
\begin{equation}
8\pi p=-\frac{2R_{t\,t}}{R}-\frac{R_{t}^{2}}{R^{2}}-\frac{\sigma_{t}R}{R_{t}%
}-3\sigma\label{p9}%
\end{equation}

Clearly in the particular case when $\sigma=k/R^{2}$ then (\ref{den19}) and
(\ref{p9}) give rise to the Friedmann equations. Note also that on writing%
\begin{equation}
\sigma=\bar{\sigma}+\frac{\chi}{3} \label{sig34}%
\end{equation}
where $\bar{\sigma}=\bar{\sigma}\left(  t\right)  $ and $\chi$ is a constant
then (\ref{den19}) and (\ref{p9}) become respectively:%
\begin{equation}
8\pi\rho=3\left(  \frac{R_{t}^{2}}{R^{2}}+\bar{\sigma}\right)  +\chi
\qquad\qquad8\pi p=-\frac{2R_{t\,t}}{R}-\frac{R_{t}^{2}}{R^{2}}-\frac
{\bar{\sigma}_{t}R}{R_{t}}-3\bar{\sigma}-\chi\label{combo34}%
\end{equation}
Thus the choice of $\sigma$ at (\ref{sig34}) results in the systematic
inclusion of a constant, $\chi$ equivalent to the cosmological constant
$\Lambda$ ($\equiv-\chi)$ normally associated with Einstein's equations.
However, in terms of cosmological applications it is better to consider
equations (\ref{den19}) and (\ref{p9}) in their full generality as $\sigma$ is
\textit{a priori} an undefined function of $t$ and so has the capacity to act
as a `variable cosmological constant'. For example, Perivolaropoulos
\cite{peri} describes the \textit{cosmological constant problem} whereby the
cosmological constant is considered to have had relatively large value during
the early period of rapid inflation much larger than during the current epoch.

Finally note that the metric (\ref{metric}) with (\ref{mu}) and (\ref{lam})
remains invariant under the transformation%
\begin{equation}
r=\frac{1}{\bar{r}}\qquad\qquad R=\frac{1}{\sigma\bar{R}} \label{trans}%
\end{equation}
where $\bar{R}=\bar{R}\left(  t\right)  $. In particular the transformation
gives to%
\begin{equation}
ds^{2}=e^{2\bar{\lambda}}dt^{2}-e^{2\bar{\mu}}\left(  d\bar{r}^{2}+\bar{r}%
^{2}d\Omega^{2}\right)  \label{metric5}%
\end{equation}
where%
\begin{equation}
e^{-\bar{\mu}}=\dfrac{\left(  1+\dfrac{\sigma\bar{R}^{2}r^{2}}{4}\right)
}{\bar{R}}\qquad\qquad e^{\bar{\lambda}}=\left(  \frac{\bar{R}_{t}}{\bar{R}%
}-\frac{r^{2}\bar{R}\left(  2\sigma\bar{R}_{t}+\sigma_{t}\bar{R}\right)
}{4\left(  1+\dfrac{\sigma\bar{R}^{2}r^{2}}{4}\right)  }\right)  a^{-\frac
{1}{2}} \label{mu5}%
\end{equation}

Thus the properties of the energy density and pressure found at $\left(
r,t\right)  $ are replicated at $\left(  \bar{r},t\right)  $ using
(\ref{trans}). In particular the properties found at $r=0$ are replicated at infinity.

\section{Vacuum energy and equation of state $p=-\rho$}

Consider now the case in (\ref{sig34}) when%
\begin{equation}
\bar{\sigma}=-\dfrac{R_{t}^{2}}{R^{2}}\qquad\qquad\sigma=\bar{\sigma}%
+\frac{\chi}{3} \label{sigbar}%
\end{equation}
so that equation (\ref{combo34}) reduces to the vacuum energy equation of
state (see for example, Peacock \cite{peacock})%
\begin{equation}
8\pi\rho=\chi\qquad\qquad p=-\rho\label{combo36}%
\end{equation}
The full solution is now given by equations (\ref{mu}) and (\ref{lam})%
\begin{equation}
e^{-\mu}=\frac{1}{R}\left\{  1+\frac{r^{2}R^{2}}{4}\left(  \dfrac{\chi}%
{3}-\dfrac{R_{t}^{2}}{R^{2}}\right)  \right\}  \qquad e^{\lambda}%
=1-\frac{r^{2}R^{2}\left(  \dfrac{\chi}{3}-\dfrac{R_{tt}}{R}\right)
}{2\left\{  1+r^{2}R^{2}\left(  \dfrac{\chi}{3}-\dfrac{R_{t}^{2}}{R^{2}%
}\right)  \right\}  } \label{soln5}%
\end{equation}
If proper time is employed then on writing $\alpha^{2}=\chi/3$ with
(\ref{sigbar}) then%
\begin{equation}
R=e^{\alpha t}\qquad\qquad e^{-\mu}=\frac{1}{R}\qquad\qquad e^{\lambda
}=1\qquad\qquad\label{soln11}%
\end{equation}
and also%
\begin{equation}
\sigma=0\qquad\qquad m=\frac{cr^{3}e^{3\alpha t}}{2} \label{soln11a}%
\end{equation}
When $\chi=0$ then from (\ref{mass5}) a Minkowski spacetime is obtained
represented by%
\begin{equation}
e^{-\mu}=\frac{1}{R}\left\{  1-\frac{r^{2}R_{t}^{2}}{4}\right\}  \qquad\qquad
e^{\lambda}=1+\frac{r^{2}RR_{tt}}{2\left\{  1-\dfrac{r^{2}R_{t}^{2}}%
{4}\right\}  } \label{czero}%
\end{equation}
Thus using proper time%
\begin{equation}
R=1+\alpha t\qquad\qquad e^{-\mu}=\frac{1}{R}\left\{  1-\frac{\alpha^{2}r^{2}%
}{4}\right\}  \qquad\qquad e^{\lambda}=1 \label{vac5}%
\end{equation}
with%
\begin{equation}
\sigma=-\frac{\alpha^{2}}{\left(  1+\alpha t\right)  ^{2}}\qquad\qquad m=0
\label{vac7}%
\end{equation}
Equation (\ref{vac5}) is of a form described by Milne \cite{milne} and Peacock
\cite{peacock}.

\section{Equations of state}

\subsection{Cosmological application}

Consider the cosmological case so that there is no fluid boundary and where it
will be assumed that a fluid equation of state $p=p\left(  \rho\right)  $
exists at $r=0$. In addition local proper time is used so that so that from
equation (\ref{k21})%
\begin{equation}
a=\frac{R_{t}^{2}}{R^{2}} \label{prop12}%
\end{equation}
With $K=R,$ equation (\ref{cos10}) may be solved for particular $p=p\left(
\rho\right)  $\ and the results which have a familiar form are summarised in
Table 1.

\begin{center}
\bigskip%
\begin{tabular}
[c]{|c|c|c|}\hline%
\begin{tabular}
[c]{l}%
\textbf{Equation of state}\\
\multicolumn{1}{c}{$\left(  \text{ at }r=0\right)  $}%
\end{tabular}
& \textbf{Pressure }$p$\textbf{ } &
\begin{tabular}
[c]{l}%
\textbf{Energy density }$\rho$\\
\multicolumn{1}{c}{$\left(  \alpha\text{ is constant}\right)  $}%
\end{tabular}
\\\hline\hline%
\begin{tabular}
[c]{l}%
Adiabatic\\
\multicolumn{1}{c}{$\left(  \gamma\neq1\right)  $}%
\end{tabular}
& $p=N\rho^{\gamma}$ & $\rho=\left(  \alpha R^{3\left(  \gamma-1\right)
}-N\right)  ^{\dfrac{1}{1-\gamma}}$\\\hline%
\begin{tabular}
[c]{l}%
Linear general\\
$\left(  \text{all values of n}\right)  $%
\end{tabular}
& $p=(n-1)\rho$ & $\rho=\dfrac{\alpha}{R^{3n}}$\\\hline%
\begin{tabular}
[c]{c}%
Dust\\
\multicolumn{1}{l}{$\left(  n=1\right)  $}%
\end{tabular}
& $p=0$ & $\rho=\dfrac{\alpha}{R^{3}}$\\\hline
$\
\begin{tabular}
[c]{c}%
Radiation domination\\
$\left(  n=4/3\right)  $%
\end{tabular}
\ \ \ $ & $p=\dfrac{\rho}{3}$ & $\rho=\dfrac{\alpha}{R^{4}}$\\\hline%
\begin{tabular}
[c]{c}%
Vacuum energy\\
$\left(  n=0\right)  $%
\end{tabular}
& $p=-\rho$ & $\rho=\alpha$\\\hline
\end{tabular}

\smallskip

\textbf{Table 1}

\bigskip
\end{center}

Thus the function $\sigma\left(  t\right)  $ can then be determined from
equation (\ref{den19}) so that%
\begin{equation}
\sigma=\frac{8\pi\rho}{3}-\frac{R_{t}^{2}}{R^{2}} \label{sig43}%
\end{equation}
This equation determines the function $\sigma\left(  t\right)  $. In FRW
cosmology $\sigma=k/R^{2}$ then equation (\ref{sig43}) is a differential
equation that is solved to determine $R=R\left(  t\right)  $.

\subsection{ Gravitational collapse application}

In cases of gravitational collapse suppose that a fluid sphere has an equation
of state $p=p\left(  \rho\right)  $ defined at the centre $r=0$ which from
equation (\ref{state}) gives results which are identical to those described in
Table 1. In this context the adiabatic equation of state is particularly
relevant to polytropic stars as has been described by Weinberg \cite{wein}.
However it is not assumed that local proper time is used at $r=0$ \ and so
equation (\ref{k21}) does not hold. However it is supposed that the fluid body
has a well defined boundary $r=b$ which matches the Scharzschild vacuum
solution and where the boundary value of the pressure must be zero.

For zero pressure at $r=b$ it is well known, see for example, Bonnor \&
Vickers \cite{bon} or Cahill and McVittie \cite{ca} that the mass function
(\ref{mass5}) is a constant $m=M.$ Thus%
\begin{equation}
M=\frac{4\pi\rho S^{3}}{3}=\frac{4\pi\rho B^{3}}{3} \label{mass21}%
\end{equation}
where%
\begin{equation}
B=B\left(  t\right)  =\frac{Rb}{1+\dfrac{\sigma R^{2}b^{2}}{4}} \label{b5}%
\end{equation}
so that%
\begin{equation}
\frac{d}{dt}\left(  \rho B^{3}\right)  =0\qquad\Longrightarrow\qquad
B=\frac{\kappa}{\rho^{1/3}} \label{b10}%
\end{equation}
Equations (\ref{b5}) together with (\ref{b10}) may used to show that:%
\begin{equation}
\sigma=-\frac{4}{R^{2}b^{2}\kappa}\left(  \kappa+Rb\rho^{1/3}\right)
\label{sig27}%
\end{equation}
whilst from equation (\ref{rho5})%
\begin{equation}
a=\frac{8\pi\rho}{3}+\frac{4}{R^{2}b^{2}\kappa}\left(  \kappa+Rb\rho
^{1/3}\right)  \label{a27}%
\end{equation}
\ It follows for (\ref{state}) with (\ref{b10}) that%
\begin{equation}
p\frac{\partial S^{3}}{\partial t}=-\frac{\partial\left(  \rho\left[
S^{3}-B^{3}\right]  \right)  }{\partial t} \label{press21}%
\end{equation}
which gives the expression for pressure such that $p\left(  b,t\right)  =0$ as required.

Finally note that each of these fluid spheres when endowed with first order
rotation in terms of a rotation parameter may also be matched to empty
space-time using a general result given by Wiltshire \cite{wiltshire2}.

\section{Hubble constant and deceleration parameter}

A further understanding of the geometry of this solution and its relation to
current cosmological observational data may be obtained by determining the
Hubble constant and deceleration parameter for the solution.

The null geodesic equation for an inward travelling photon as described by an
observer at $\left(  r,t\right)  =\left(  0,t_{0}\right)  $ using proper time
is%
\begin{equation}
\frac{dr}{dt}=-e^{\lambda-\mu}\qquad\text{with}\qquad e^{\lambda
}=1\label{null5}%
\end{equation}
With the notation that%
\begin{equation}
t=t_{0}-\tau\label{t5}%
\end{equation}
where $\tau$ is the travel time of the photon and using (\ref{mu}) and
(\ref{lam}) equation (\ref{null5}) expressed in terms of travel time $\tau$
becomes:%
\begin{equation}
\frac{dr}{d\tau}=\frac{1+r^{2}\sigma R^{2}}{R}\qquad\qquad e^{\lambda}=\left(
\frac{R_{t}}{R}-\frac{r^{2}R\left(  2\sigma R_{t}+\sigma_{t}R\right)
}{4\left(  1+\dfrac{\sigma R^{2}r^{2}}{4}\right)  }\right)  a^{-\frac{1}{2}%
}=1\label{check5}%
\end{equation}
The equations (\ref{check5}) are solved upto and including second order terms
in $\tau$ by writing%
\begin{align}
R &  =R_{0}-\dot{R}_{0}\tau+\frac{\ddot{R}_{0}}{2}\tau^{2}\qquad\qquad
\sigma=\sigma_{0}-\dot{\sigma}_{0}\tau+\frac{\ddot{\sigma}_{0}}{2}\tau
^{2}\nonumber\\
a &  =a_{0}-\dot{a}_{0}\tau+\frac{\ddot{a}_{0}}{2}\tau^{2}\label{expan6}%
\end{align}
where for example $\dot{R}$ means $R_{t}$ evaluated at $t=t_{0}$. In this way
the solution the first of (\ref{check5}) and for example the redshift $z$ can
be calculated in the usual way%
\begin{equation}
r\left(  \tau\right)  R_{0}=\tau+\frac{H}{2}\tau^{2}\qquad\text{and \qquad
}z=H\tau+H^{2}\frac{\left(  q+2\right)  }{2}\tau^{2}\label{null20}%
\end{equation}
where the Hubble constant and deceleration parameter are%
\begin{equation}
H=\frac{\dot{R}_{0}}{R_{0}}\qquad\qquad\qquad q=-\frac{R_{0}\ddot{R}_{0}}%
{\dot{R}_{0}^{2}}\label{param67}%
\end{equation}
The condition that $e^{\lambda}=1$ is given by%
\begin{equation}
a_{0}=\frac{\dot{R}_{0}^{2}}{R_{0}^{2}}\qquad\qquad\qquad\dot{a}_{0}%
=\frac{2\dot{R}_{0}\ddot{R}_{0}}{R_{0}^{2}}-2\frac{\dot{R}_{0}^{3}}{R_{0}^{3}%
}\label{a20}%
\end{equation}
and is essentially the statement that $a=\left(  R_{\,t}/R\right)  ^{2}$ upto
and including first order terms in $\tau$.

With this notation the first of (\ref{a20}) with (\ref{rho5}) and
(\ref{param67}) gives%
\begin{equation}
H^{2}=\frac{\dot{R}_{0}^{2}}{R_{0}^{2}}=\frac{8\pi\rho_{0}}{3}-\sigma_{0}
\label{hub20}%
\end{equation}
whilst the second of (\ref{a20}) with (\ref{param67}) together with
(\ref{state5}) results in%
\begin{equation}
q=-\frac{R_{0}\ddot{R}_{0}}{\dot{R}_{0}^{2}}=\frac{\dot{\sigma}_{0}}{2\left(
\dfrac{8\pi\rho_{0}}{3}-\sigma_{0}\right)  ^{3/2}}+\frac{4\pi\left(
p_{0}+\rho_{0}\right)  }{\left(  \dfrac{8\pi\rho_{0}}{3}-\sigma_{0}\right)
}-1 \label{qacc20}%
\end{equation}
So when $p=(n-1)\rho$ then equations (\ref{hub20}) and (\ref{qacc20}) may be
taken to give%
\begin{equation}
q=\frac{\dot{\sigma}_{0}}{2\left(  \dfrac{8\pi\rho_{0}}{3}-\sigma_{0}\right)
^{3/2}}+\frac{4\pi n\rho_{0}}{\left(  \dfrac{8\pi\rho_{0}}{3}-\sigma
_{0}\right)  }-1 \label{qresult}%
\end{equation}
Thus in the case when $\sigma=0$ for all $t$ then%
\begin{equation}
q=\frac{3n}{2}-1 \label{qol}%
\end{equation}
as expected. In addition for the case of vacuum energy $n=0$ and $\sigma$ is
constant then $q=-1$ again as required.

In the light of recent observations for example WMAP \cite{WM} evidence from
the cosmic microwave background suggests that the universe is essentially flat
so that\ in the context of this model $\sigma_{0}=0$ so that%
\begin{equation}
H^{2}=\frac{8\pi\rho_{0}}{3} \label{hubex}%
\end{equation}
as is normally calculated. In addition however observations of distant type Ia
supernovae suggest that $q<0$ and that the universe is decelerating rapidly.
In this model (\ref{qresult}) can be rearranged for $\dot{\sigma}_{0}$
incorporating $\sigma_{0}=0$ so that:%
\begin{equation}
\dot{\sigma}_{0}=\left(  \frac{8\pi\rho_{0}}{3}\right)  ^{3/2}\left(
2q-3n+2\right)  \label{sigd}%
\end{equation}
\ With equation (\ref{hubex}) this is also%
\begin{equation}
\dot{\sigma}_{0}=H^{3}\left(  2q-3n+2\right)  \label{sige}%
\end{equation}
Thus for example a matter dominated universe, $n=1$ with $q=-1/2$ gives%
\begin{equation}
\dot{\sigma}_{0}=-2H^{3} \label{sigex}%
\end{equation}
Hence it has be shown how the observed data can be used to determine the
geometric parameters $\left[  \sigma_{0}\text{, }\dot{\sigma}_{0}\right]  $
which characterise the difference between the geometry of the current model
with that of standard FRW cosmology.

\section{Conclusion}

In this paper the primary focus has been on consideration of an extended
version of the FRW solution which can be characterised by a Ricci curvature
scalar having a value $6\sigma\left(  t\right)  $ which is independent of the
scale factor $R\left(  t\right)  $. Moreover the purely time dependent energy
density consists of the sum of two terms which from a geometrically point of
view are the curvature and also a function $a\left(  t\right)  $ that defines
the nature of the time coordinate, for example proper time. This is an
extended form of the Friedmann equation and contrasts with the FRW solution
for which the curvature and also the energy density depend closely on the
scale factor. It has further been shown how expression for internal pressure
now a function of $\left(  r,t\right)  $ gives rise to an equation which is
closely analogous to the second Friedmann equation which expresses the
acceleration of the scale factor in the standard model. The generalised
equations are then shown by a translation of the curvature function to contain
naturally a constant that can be interpreted as a cosmological constant and
can also be used to describe vacuum energy. However its introduction is
unnecessary in the new model as it is better to to consider the curvature in
its full generality for the purposes of cosmological application. For these
purposes it is shown how to define local proper time in such a way that
observer essentially see a homogeneous cosmology but with the inclusion of the
modified curvature term. The Hubble constant and deceleration parameter are
calculated in the usual way but now include a component that reflects the
generalised form of the curvature term. Hence when the observational data (an
essentially flat universe coupled with a negative deceleration parameter) is
introduced the Hubble constant and deceleration parameter are shown to provide
information about the generalised nature of the curvature and the geometry of
the modified cosmological model. In terms of this model the data is not
interpreted in terms of dark energy or cold dark matter.

Finally the solution is discussed using a equation of state for an observer
employing local proper time. Moreover the resulting solutions are also
described in terms of problem of gravitational collapse. In particular it is
shown how fluids including  polytropes, with a central equation of state may
be matched to Schwarzschild empty space-time.

\end{document}